\documentclass[12pt]{article}

\usepackage{graphicx}
\usepackage{amsfonts}
\def\+{{+\!\!\!+}} 
\def\pp{\mbox{\tiny${}_{\stackrel\+ =}$}}

\def\d{\partial}

\def\P{\Phi}

\def\p{\psi} 
 
\def\e{\varepsilon}

\def\pmb#1{\setbox0=\hbox{#1}%
\kern.0em\copy0\kern-\wd0 
\kern-.04em\copy0\kern-\wd0 
\kern.08em\copy0\kern-\wd0 
\kern-.04em\raise.0433em\box0 }         
 
\newcommand{\nc}{\newcommand} 
\nc{\beq}{\begin{equation}} 
\nc{\eeq}[1]{\label{#1}\end{equation}} 
\nc{\ber}{\begin{eqnarray}} 
\nc{\eer}[1]{\label{#1}\end{eqnarray}} 
\nc{\pek}[1]{\cite{#1}} 
\nc{\enr}[1]{(\ref{#1})} 
\nc{\kal}[1]{{\cal{#1}}} 
\nc{\dott}{\;\cdot\;} 
\newcommand{\Section}[1]{\section{#1} \setcounter{equation}{0}}

\def\0 {\nonumber}

\begin{document} 
\setcounter{page}{0}
\newcommand{\inv}[1]{{#1}^{-1}} 
\renewcommand{\theequation}{\thesection.\arabic{equation}} 
\newcommand{\be}{\begin{equation}} 
\newcommand{\ee}{\end{equation}} 
\newcommand{\bea}{\begin{eqnarray}} 
\newcommand{\eea}{\end{eqnarray}} 
\newcommand{\re}[1]{(\ref{#1})} 
\newcommand{\qv}{\quad ,} 
\newcommand{\qp}{\quad .} 

\thispagestyle{empty}
\begin{flushright} \small
UUITP-01/06 \\ HIP-2006-11/TH \\ 
\end{flushright}
\smallskip
\begin{center} \LARGE
{\bf Generalized K\"ahler Geometry from supersymmetric sigma models}
 \\[12mm] \normalsize
{\bf Andreas Bredthauer$^{a}$, Ulf~Lindstr\"om$^{a,b}$,  Jonas Persson$^a$, and Maxim Zabzine$^{a}$} \\[8mm]
 {\small\it
$^a$Department of Theoretical Physics 
Uppsala University, \\ Box 803, SE-751 08 Uppsala, Sweden \\
~\\
$^b$HIP-Helsinki Institute of Physics, University of Helsinki,\\
P.O. Box 64 FIN-00014  Suomi-Finland\\
~\\
}
\end{center}
\vspace{10mm}
\centerline{\bfseries Abstract} \bigskip
 We give a physical derivation of generalized K\"ahler geometry. Starting from a supersymmetric nonlinear 
 sigma model, we rederive  and explain the results of Gualtieri \cite{gualtieri} regarding the equivalence between generalized 
 K\"ahler geometry and the bi-hermitean geometry of Gates-Hull-Ro\v{c}ek \cite{Gates:1984nk}. 
 When cast in the language of supersymmetric  sigma models, this relation maps precisely to that
 between the Lagrangian and the Hamiltonian formalisms.\\
 We also discuss topological twist in this context.

\noindent  

\eject


\section{Introduction}

 The intimate relation between supersymmetry and complex geometry has long been known. It is, e.g.,  more 
than twenty-five years since the target-space geometry of supersymmetric nonlinear sigma models was first classified
\cite{Zumino:1979et}, \cite{Alvarez-Gaume:1980vs}, and for two dimensional models, where the    
 background geometry 
may include an antisymmetric $B-$field, the classification dates over twenty years back \cite{Gates:1984nk}. It is thus quite 
remarkable that the appropriate mathematical setting for describing some of the target space geometries  has only arisen in a very recent development \cite{Hitchin},
\cite{gualtieri}. To be more precise, in \cite{Gates:1984nk} it is shown that a $2d$ supersymmetric nonlinear sigma model 
formulated in $N=(1,1)$ superspace have a second (non-manifest) supersymmetry when the target-space has a certain 
bi-hermitean geometry. We refer to this as Gates-Hull-Ro\v cek geometry (GHRG). 
Apart from the metric and $B$-field, the geometry also supports two complex structures and the metric is hermitean with 
respect to both. In \cite{gualtieri} it is now shown that  these geometric objects and their transformations have a unified 
description as  
Generalized K\"ahler Geometry (GKG). This is a geometry where the fundamental geometric objects live on the sum of the tangent 
and the cotangent spaces. It is a subset of Generalized Complex Geometry (GCG).

The map between the bi-hermitean geometry of  \cite{Gates:1984nk} and GKG is somewhat 
complicated and also independent of the sigma model realization. This begs the question of whether there
is a generalized sigma model which directly realizes GKG, i.e., where an additional supersymmetry exists iff the 
target space has GKG. Such a realization would also shed light on the question of when the second supersymmetry closes off-shell, 
a question left open in \cite{Gates:1984nk}.

Such generalized sigma model, with fields transforming also in the cotangent space, was constructed in \cite{Lindstrom:2004eh} and studied
from the GKG point of view in  \cite{Lindstrom:2004iw}, \cite{Bergamin:2004sk},\cite{Bredthauer:2005zx}. For certain models with less supersymmetry
a direct relation to GCG was found. Due to the large number of possible
first order Lagrangians and the complexity of the problem, the general case was not completely sorted out.

At the same time, there was an open question regarding the manifest $N=(2,2)$ formulation of the sigma models. 
The question here is if the known
$N=(2,2)$ superfields, the chiral, twisted chiral and semi-chiral fields, are sufficient when formulating the most general
$N=(2,2)$ sigma model.   Using information gained from the $N=(1,1)$ analysis, the answer to this question was recently shown to be yes: 
Any GKG may be coordinatized by  (anti-)chiral, twisted (anti-)chiral and semi (anti-)chiral\footnote{This $N=(2,2$) representation was introduced in \cite{Buscher:1987uw}.} superfields and further
has a potential $K$ in terms of which both the metric and the $B$-field may be described  \cite{Lindstrom:2005zr}.

In view of the above, a fairly complete picture of the target space of $N=(2,2)$ supersymmetric sigma models has emerged:
The target space geometry is GKG. This geometry can always be coordinatized by the three kinds of $N=(2,2)$ superfields mentioned, 
and the $N=(2,2)$ sigma model action is the potential $K$ (which actually has the geometric meaning of a generating function for 
symplectomorphisms). In terms of  $N=(1,1)$ {\em scalar} fields, the second supersymmetry algebra may or may not close off-shell 
depending on whether the commutator of the two complex structures vanishes or not. The underlying geometry is always GKG, but 
the relation to the sigma model target space is not straight forward. We may add auxiliary {\em spinor} superfields to get a generalized $N=(1,1)$ sigma model and have the 
second supersymmetry close off-shell and to obtain a target space that has a possibility of being more directly related  
to GKG. The spinor fields, transforming as target space one-forms, may be added in a way which corresponds to a dimensional reduction of the 
$N=(2,2)$ sigma model, in which case the relation to GKG is clear. There are many other possible ways of adding auxiliary fields for closure, 
however, and in the Lagrangian formalism we do not seem  to be uniquely led to GKG.

The remaining question is thus: Starting from a $N=(1,1)$ of the sigma model, is there a formulation of this model where a closing 
second supersymmetry uniquely leads to GKG?

A clue to this question was given in \cite{Zabzine:2005qf} in which the present article 
takes its starting point. There the Hamiltonian formulation of $N=(1,1)$ supersymmetric sigma models  is discussed in general terms
and a relation to GCG is derived. The discussion is model-independent, and to find out the precise GCG, an explicit Hamiltonian is needed.
A first application with an explicit Hamiltonian was presented in \cite{Calvo:2005ww}, where a Poisson sigma model is discussed.
In the present article we show that Gates-Hull-Ro\v{c}ek geometry and GKG are identical  in the Hamiltonian formulation of the sigma model.
 
 The paper is organized as follows. In Section \ref{N1general} we review the supersymmetric 
  version of the cotangent bundle of the loop space.
  Section \ref{N1sigma} discusses the superfield Hamiltonian formalism for $N=(1,1)$ 
  sigma model.  In Section \ref{N2general} we review the result from \cite{Zabzine:2005qf} regarding 
   the relation between the extended supersymmetry in phase space and generalized complex
    structure. Next, in Section \ref{N2sigma} we present  the phase-space realization of 
     $N=(2,2)$ sigma models. We show that its phase-space description  is naturally related 
      to GKG.  In Section \ref{GHR} the relation between 
       the Gates-Hull-Ro\v{c}ek and GK  geometrical data is presented. 
        We obtain complete agreement with Gualtieri's result \cite{gualtieri}. 
         In Section \ref{twist} we discuss the topological twist in the Hamiltonian context.
    Finally, in Section \ref{end} we summarize our results and list some open problems.

 \section{$N=1$ supersymmetry in phase space}
 \label{N1general}
 
  There is a large class of two-dimensional models that have the same phase space as the sigma model.
 This phase space is the cotangent bundle of the loop space. 
 In  this Section we review the supersymmetric version of  this construction. More details 
  can be found in  \cite{Zabzine:2005qf}.  
  
 For a world-sheet $\Sigma= S^1 \times \mathbb{R}$ the phase space may be identified with
  a cotangent bundle $T^*LM$  of the loop space $LM=\{ X: S^1 \rightarrow M \}$. Using local 
   coordinates $X^\mu(\sigma)$ and their conjugate momenta $p_\mu(\sigma)$ the usual 
    symplectic form on $T^*LM$ is given by 
  \beq
  \omega = \int\limits_{S^1} d\sigma\,\, \delta X^\mu \wedge \delta p_\mu ,
  \eeq{stanafshq200} 
    where $\delta$ is the de Rham differential on $T^*LM$. The symplectic form (\ref{stanafshq200})
   can be twisted by a closed three form $H \in \Omega^3(M)$, $dH=0$ as follows
  \beq
  \omega = \int\limits_{S^1} d\sigma\,\, \left ( \delta X^\mu \wedge \delta p_\mu
   + H_{\mu\nu\rho} \d X^\mu  \delta X^\nu \wedge \delta X^\rho \right ),
  \eeq{stanafshq200more} 
 where $\d \equiv \d_\sigma$ is derivative with respect to $\sigma$.  For the symplectic 
  structure (\ref{stanafshq200more}) the transformations
  \beq
   X^\mu \rightarrow X^\mu,\,\,\,\,\,\,\,\,\,\,\,\,\,\,\,\,\,\,p_\mu \rightarrow p_\mu + b_{\mu\nu} \d X^\nu
  \eeq{canonicalshdj89}
  are symplectic if $b$ is a closed two form, $b \in \Omega^2(M)$, $db=0$. 
  
  Next we construct the supersymmetric version of  $T^*LM$. Let $S^{1,1}$ be a 
  "supercircle" with coordinates $(\sigma, \theta)$. The corresponding superloop 
   space is defined by ${\cal L}M =\{\phi: S^{1,1} \rightarrow M\}$. The phase space 
    is given by the cotangent bundle of ${\cal L}M$ with reversed 
     parity on the fibers, denoted $\Pi T^*{\cal L}M$. In local coordinates we have a scalar superfield
       $\phi^\mu(\sigma, \theta)$
      and a conjugate spinorial superfield $S_\mu(\sigma, \theta)$ with the 
       expansion\footnote{We choose our conventions in such way that we can match it 
        with the supersymmetric sigma model.  Of course there are other possible conventions.}
\beq
 \phi^\mu (\sigma, \theta) = X^\mu(\sigma) + \theta \lambda^\mu(\sigma),\,\,\,\,\,\,\,\,\,\,\,\,\,\,\,\,\,\,\,\,
 S_\mu(\sigma, \theta) = \rho_\mu(\sigma) + i \theta p_\mu(\sigma),
\eeq{exkap[ajw9899}
   where $\lambda$ and $\rho$ are fermions. The  symplectic structure on 
    $\Pi T^* {\cal L}M$ is taken to be
   \beq
  \omega = i \int d\sigma d\theta\,\, \left ( \delta S_\mu \wedge \delta \phi^\mu - H_{\mu\nu\rho}
   D\phi^\mu\, \delta \phi^\nu \wedge \delta \phi^\rho \right ),
 \eeq{definsymlpedjsdo}
  such that the bosonic part of (\ref{definsymlpedjsdo}) coincides with (\ref{stanafshq200more}). 
   The technical details of the definition of the corresponding Poisson brackets are collected 
    in Appendix B.  We denote the Poisson bracket by $\{\,\,,\,\,\}$ when $H=0$ and  by $\{\,\,,\,\,\}_H$
     when $H\neq 0$. 
    The symplectic structure (\ref{definsymlpedjsdo}) makes $C^\infty (\Pi T^* {\cal L}M)$
     into a superPoisson algebra with the obvious $\mathbb{Z}_2$ grading. 
  
    In superspace we have two natural operations, $D$ and $Q$.
   The derivative $D$ is defined as
   \beq
     D= \frac{\d }{\d \theta} + i \theta \d
   \eeq{definsoldfkfk}
    and the operator $Q$ as
    \beq
     Q = \frac{\d}{\d \theta} - i \theta \d.
    \eeq{djkfdhe892902}
    $D$ and $Q$ satisfy the following algebra
    \beq
     D^2 = i\d,\,\,\,\,\,\,\,\,\,\,\,\,\,\,\,\,\,
      Q^2 = - i\d,\,\,\,\,\,\,\,\,\,\,\,\,\,\,\,\,\,
      DQ + QD = 0 .
    \eeq{alshjwopqp[p}
     We begin by considering the case $H=0$.
 On $C^\infty (\Pi T^*{\cal L}M)$  we introduce the generator
  \beq
   {\mathbf Q}_1 (\epsilon) = -  \int\limits_{S^{1,1}} d\sigma d\theta\,\, \epsilon S_\mu Q\phi^\mu,
  \eeq{manidkalalal122}
   where $\epsilon$ is an odd parameter. Using the symplectic structure (\ref{definsymlpedjsdo}) 
    we calculate the Poisson brackets
    \beq
     \{ {\mathbf Q}_1 (\epsilon), {\mathbf Q}_1(\tilde{\epsilon}) \} =
      {\mathbf P} (2\epsilon \tilde{\epsilon}),
    \eeq{djfkfllla;;;app292}
     where ${\mathbf P}$ is the generator of translations along $\sigma$,
     \beq
     {\mathbf P}(a) = \int\limits_{S^{1,1}} d\sigma d\theta\,\,a S_\mu\d \phi^\mu ,
          \eeq{definfoo38383830}
    with $a$ an even parameter.  ${\mathbf Q}_1 (\epsilon)$ generates the following 
     transformation on the fields
      \beq
     \delta_1(\epsilon) \phi^\mu = \{\phi^\mu, {\mathbf Q}_1 (\epsilon)\} =
     - i \epsilon Q \phi^\mu ,\,\,\,\,\,\,\,\,\,\,\,\,\,\,\,\,
     \delta_1(\epsilon) S_\mu =  \{S_\mu, {\mathbf Q}_1 (\epsilon)\} =
     - i \epsilon Q S_\mu .
     \eeq{transkd829292}
     It is natural to refer to these transformations as supersymmetry transformations. 
     
   Next we consider the case $H\neq 0$ with Poisson bracket $\{\,\,,\,\,\}_H$.
    There is a convenient technical trick to write down the expressions involving $H$.    
   Namely,  we can generate $H$ by performing a  non-canonical transformation
     from the situation with $H=0$;
    \beq
     \phi^\mu \rightarrow \phi^\mu,\,\,\,\,\,\,\,\,\,\,\,\,\,\,\,\,\,\,\,
      S_\mu \rightarrow S_\mu - B_{\mu\nu} D\phi^\nu, 
    \eeq{perifofk83999}
    where we define the field strength $H$ as
   \beq
    H_{\mu\nu\rho} = \frac{1}{2} (B_{\mu\nu,\rho} + B_{\nu\rho,\mu} + B_{\rho\mu,\nu})~.
   \eeq{definiofHH}
    Indeed all final expressions will depend only on $H$, but not on $B$.  
   Thus the supersymmetry generator has the form 
 \beq
   {\mathbf Q}_1 (\epsilon) = -  \int\limits_{S^{1,1}} d\sigma d\theta\,\, \epsilon (S_\mu - B_{\mu\nu} D\phi^\nu) Q\phi^\mu,
  \eeq{manidkalalal122new}   
   but its component expansion depends only on $H$. This generator satisfies the same 
    algebra (\ref{djfkfllla;;;app292}) as before  with $\{\,\,,\,\,\}$ replaced by $\{\,\,,\,\,\}_H$ and the generator
     of translations is given by
       \beq
     {\mathbf P}(a) = \int\limits_{S^{1,1}} d\sigma d\theta\,\,a\left (S_\mu\d \phi^\mu +
      i H_{\mu\nu\rho} D\phi^\mu Q\phi^\nu Q\phi^\rho \right ).
     \eeq{definfoo38383830new}
      Again we could obtain the above expression by performing the shift (\ref{perifofk83999})
       in the generator (\ref{definfoo38383830}). 
      The supersymmetry transformations are given by the same expressions as before
       \beq
     \delta_1(\epsilon) \phi^\mu = \{\phi^\mu, {\mathbf Q}_1 (\epsilon)\}_H =
     - i \epsilon Q \phi^\mu ,\,\,\,\,\,\,\,\,\,\,\,\,\,\,\,\,
     \delta_1(\epsilon) S_\mu =  \{S_\mu, {\mathbf Q}_1 (\epsilon)\}_H =
     - i \epsilon Q S_\mu ,
     \eeq{transkd829292new}
 but where the generator $\mathbf Q$  is now defined in  (\ref{manidkalalal122new}). In forthcoming discussions of the case 
  $H\neq 0$ we will often introduce $H$ via the non-canonical transformation trick. This 
   is merely a  calculational shortcut, however, and all results can just as well be derived from first principles.

\section{Hamiltonian treatment of the $N=(1,1)$ sigma model}
\label{N1sigma}

In this Section we give the Hamiltonian formulation of a $N=(1,1)$ sigma model using
a treatment closely related to dimensional reduction. We will derive the appropriate 
 Hamiltonian for $N=(1,1)$ sigma model written in the Hamiltonian superfields $(\phi, S)$.
 
 For the sake of clarity 
we start from the $N=(1,1)$ sigma model with $B=0$. The action of this model is 
\beq
 S = \frac{1}{2}\int d^2\sigma\,d\theta^+ d\theta^- \,\,D_+ \Phi^\mu D_-\Phi^\nu g_{\mu\nu}(\Phi),
\eeq{11actsta}
 written in  $N=(1,1)$ superfield notation (see Appendix A).
  Using the  conventions from the Appendix we introduce  a new set of 
   odd coordinates and derivatives.
We define the new odd coordinates as
\beq
 \theta^0 = \frac{1}{\sqrt{2}}(\theta^+ - i \theta^-),\,\,\,\,\,\,\,\,\,\,\,\,\,\,\,\,\,\,\,
 \theta^1 =\frac{1}{\sqrt{2}}(\theta^+ +i\theta^-)
\eeq{newthetajd}
and introduce the new spinor derivatives by
\beq
 D_0 =\frac{1}{\sqrt{2}}(D_+ + i D_-),\,\,\,\,\,\,\,\,\,\,\,\,\,\,\,\,\,\,\,
 D_1 =\frac{1}{\sqrt{2}}(D_+-iD_-).
\eeq{definitionneqwka}
  Explicitly $D_0$ and $D_1$ have the form
\beq
 D_0 = \frac{\d}{\d\theta^0} + i\theta^1 \d_0 + i \theta^0 \d_1,\,\,\,\,\,\,\,\,\,\,\,\,\,\,\,\,\,\,\,\,\,\,\,
 D_1 = \frac{\d}{\d\theta^1} + i\theta^0 \d_0 + i \theta^1 \d_1,
\eeq{explneders1}
and satisfy the  algebra
\beq
 D_0^2 = i \d_1,\,\,\,\,\,\,\,\,\,\,\,\,\,
 D_1^2 = i \d_1,\,\,\,\,\,\,\,\,\,\,\,\,\,
 \{ D_0, D_1\}=2i \d_0 .
\eeq{algeonewo1}
Up to boundary terms the action (\ref{11actsta}) may now be rewritten as
\beq
 S=- \frac{1}{2} \int d^2\sigma\,d\theta^1 d\theta^0\,\, D_0 \Phi^\mu D_1 \Phi^\nu g_{\mu\nu}
 = -\frac{1}{2}\int d^2\sigma\,d\theta^1 D_0(D_0 \Phi^\mu D_1\Phi^\nu g_{\mu\nu})|_{\theta^0=0}.
\eeq{newactsdl1}
 We want to integrate out $\theta^0$. To this end introduce the new superfields
 \beq
 \phi^\mu = \Phi^\mu|_{\theta^0=0},\,\,\,\,\,\,\,\,\,\,\,\,\,\,\,\,\,\,\,\,\,\,\,\,\,
  S_\mu = g_{\mu\nu} D_0 \Phi^\nu|_{\theta^0=0}
 \eeq{newsu3902}
and define  $D\equiv D_1|_{\theta^0=0}$, $\d\equiv \d_1$ and $\theta\equiv \theta^1$.
 Thus we find
\beq
 S = \int d^2\sigma\,d\theta\,\,\left (i S_\mu \d_0 \phi^\mu  - \frac{i}{2} \d_1 \phi^\mu D \phi^\nu g_{\mu\nu}
  - \frac{1}{2} S_\mu D S_\nu g^{\mu\nu} - \frac{1}{2} S_\sigma D \phi^\nu S_\gamma
   g^{\lambda\gamma} \Gamma^\sigma_{\,\,\,\,\nu\lambda} \right )~,
\eeq{actionred}
where $\Gamma^\sigma_{\,\,\,\,\nu\lambda}$ denotes the Levi-Civita connection for the metric $g_{\mu\nu}$.  From the first term we read off 
the Liouville form $\Theta$
\beq
 \Theta = i \int d\sigma d\theta\,\, S_\mu \delta \phi^\mu
\eeq{indelivsis999}
 with $\delta$ being the de Rham differential on $\Pi T^* {\cal L}M$. The symplectic 
  structure is therefore given by $\omega =\delta \Theta$ which coincides with (\ref{definsymlpedjsdo}) when $H=0$.
From (\ref{actionred})  we also read off the Hamiltonian
\beq
 {\cal H} = \frac{1}{2} \int d\sigma\,d\theta\,\,\left ( i \d\phi^\mu D\phi^\nu g_{\mu\nu}
 + S_\mu D S_\nu g^{\mu\nu} + S_\sigma D\phi^\nu S_\gamma g^{\lambda\gamma}
  \Gamma^\sigma_{\,\,\,\,\nu\lambda} \right) . 
\eeq{defhamilt}
 We have thus derived the symplectic structure
   (\ref{definsymlpedjsdo}) and the Hamiltonian (\ref{defhamilt})
  corresponding to the  $N=(1,1)$ sigma model. 

Now we have to discuss the symmetries of the Hamiltonian theory, in particular 
 supersymmetries. The original theory (\ref{11actsta}) is manifestly invariant 
  under the $N=(1,1)$ supersymmetry transformations (\ref{tfs}) since it is written 
   in $N=(1,1)$ superfields.  
 Integrating out $\theta^0$ and switching to Hamiltonian $N=1$ superfields
 have resulted in the Hamiltonian \enr{defhamilt} having two kinds of supersymmetries; 
  manifest and non-manifest ones.  We find them by reducing
 the  transformations (\ref{tfs}).  Introducing 
   \beq
    Q_0 = \frac{1}{\sqrt{2}}( Q_+ + i Q_-),\,\,\,\,\,\,\,\,\,\,\,\,\,\,\,\,\,\,
    Q_1 = \frac{1}{\sqrt{2}}(Q_+ - i Q_-)
   \eeq{newsusywe9299}
    and 
    \beq 
     \epsilon^0 = \frac{1}{\sqrt{2}}(\epsilon^+ - i\epsilon^-),\,\,\,\,\,\,\,\,\,\,\,\,\,\,\,\,\,
     \epsilon^1 = \frac{1}{\sqrt{2}}(\epsilon^+ + i \epsilon^-) 
    \eeq{newepsie2383490}
     we rewrite the transformations (\ref{tfs}) as
     \beq
      \delta \Phi^\mu = - i \epsilon^0 Q_0 \Phi^\mu - i\epsilon^1 Q_1\Phi^\mu, 
     \eeq{newsawjasiio9}
    (where the transformation rules are applied to the particular superfield $\Phi^\mu$).  Upon reduction 
     to the Hamiltonian superfields (\ref{newsu3902}) the second term on the right hand 
      side of (\ref{newsawjasiio9}) gives rise to the manifest supersymmetry (\ref{transkd829292})
       with the following identifications, $Q \equiv Q_1|_{\theta^0=0}$ and $\epsilon \equiv \epsilon^1$. 
    Furthermore the first term on the right hand side of (\ref{newsawjasiio9}) gives rise
   to the non-manifest  transformations 
\ber
\label{nonmanN1a}&&\tilde{\delta}_1(\epsilon)\phi^\mu = \epsilon g^{\mu\nu} S_\nu , \\
\label{nonmanN1b}&&\tilde{\delta}_1(\epsilon) S_\mu = i\epsilon g_{\mu\nu} \d \phi^\nu + \epsilon S_\lambda S_\sigma
g^{\lambda\rho} \Gamma^\sigma_{\,\,\,\,\mu\rho} ,
\eer{susyN1no2}
 where we have dropped the terms with $\d_0 \phi^\mu$ and $\d_0 S_\mu$.  These terms can 
  be dropped since they are generated by the Hamiltonian and correspond to time evolution. 
   Indeed one can check that the transformations (\ref{nonmanN1a})-(\ref{nonmanN1b}) 
    represents a supersymmetry.
   They satisfy the correct algebra and they leave the Hamiltonian (\ref{defhamilt}) invariant.

As mentioned in the beginning of this section, the sigma model \enr{11actsta} does not correspond to the most general background.
The general form is
\beq
 S = \frac{1}{2}\int d^2\sigma\,d\theta^+ d\theta^- \,\,D_+ \Phi^\mu D_-\Phi^\nu (g_{\mu\nu}(\Phi) +
  B_{\mu\nu}(\Phi))~,
\eeq{fullactionwthB}
and the procedure described above leads to
\beq
 S = \int d^2\sigma\,d\theta\,\,\left (i (S_\mu - B_{\mu\nu} D\phi^\nu) \d_0 \phi^\mu  - {\cal H} \right ),
 \eeq{hamsle03440}
 where the first term gives rise to  
 the Liouville form $\Theta$
\beq
 \Theta = i \int d\sigma d\theta\,\, (S_\mu - B_{\mu\nu} D\phi^\nu) \delta \phi^\mu,
\eeq{indelivsis999more}
  such that symplectic structure $\omega = \delta \Theta$ is given by (\ref{definsymlpedjsdo}).
  The second term in (\ref{hamsle03440}) corresponds to the
 Hamiltonian  
$$  {\cal H} = \frac{1}{2} \int d\sigma\,d\theta\,\,\left ( i \d\phi^\mu D\phi^\nu g_{\mu\nu}
 + S_\mu D S_\nu g^{\mu\nu} + S_\sigma D\phi^\nu S_\gamma g^{\lambda\gamma}
  \Gamma^\sigma_{\,\,\,\,\nu\lambda}  - \right .$$
  \beq
\left .  - \frac{1}{3} H^{\mu\nu\rho} S_\mu S_\nu S_\rho 
   + D\phi^\mu D\phi^\nu S_\rho H_{\mu\nu}^{\,\,\,\,\,\,\rho}  \right),
  \eeq{fullhamwithevetey}
   where we use (\ref{definiofHH}).
   The Hamiltonian is now invariant under the manifest supersymmetry  (\ref{transkd829292new})
    and under the non-manifest supersymmetry 
    which is the same as before,  \enr{nonmanN1a}-\enr{nonmanN1b}.

We would like to stress that the transformations
\beq
 \phi^\mu \rightarrow \phi^\mu,\,\,\,\,\,\,\,\,\,\,\,\,\,\,\,\,\,\,\,\,\,\,
 S_\mu \rightarrow S_\mu - b_{\mu\nu} D\phi^\nu
\eeq{acanonsj3289}
{ \em are} canonical for the special case when $b$ is a closed two form. The whole formalism is thus covariant under
 such a transformation.

\section{Extended supersymmetry  in phase space}
\label{N2general}

In this Section we review the relation between extended supersymmetry and generalized complex geometry
presented in \cite{Zabzine:2005qf}. The discussion is model-independent, i.e., independent of the particular 
choice of Hamiltonian. 

 First we consider the case with $H=0$.  In Section \ref{N1general} we described 
  the generator-algebra  (\ref{djfkfllla;;;app292}) of manifest supersymmetry on $\Pi T^* {\cal L}M$.
   Now we would like to describe the conditions under which extended supersymmetry 
    can be introduced on $\Pi T^*{\cal L}M$. A second supersymmetry should be 
     generated by some ${\mathbf Q}_2(\epsilon)$ such that it satisfies the following 
      brackets
    \beq
    \{ {\mathbf Q}_1 (\epsilon), {\mathbf Q}_2(\tilde{\epsilon}) \}=0,\,\,\,\,\,\,\,\,\,\,\,\,\,\,\,\,\,\,\,\,\,
     \{ {\mathbf Q}_2 (\epsilon), {\mathbf Q}_2(\tilde{\epsilon}) \} =
      {\mathbf P} (2\epsilon \tilde{\epsilon}).
    \eeq{djfkfllla;;;app292more}
 By dimensional arguments there is a unique ansatz for the generator ${\mathbf Q}_2(\epsilon)$
  on $\Pi T^* {\cal L} M$ which does not involve any dimensionful parameters
\beq
{\mathbf Q}_2 (\epsilon) = -  \frac{1}{2} \int d\sigma d\theta \epsilon 
 \left ( 2 D\phi^\mu S_\nu J^\nu_{\,\,\,\mu}
 + D\phi^\mu D\phi^\nu L_{\mu\nu} + S_{\mu} S_{\nu} P^{\mu\nu} \right ). 
\eeq{definsdo339390}
 We can combine $D\phi$ and $S$ into a single object
 \beq
  \Lambda = \left ( \begin{array}{c}
                        D\phi \\
                         S
                         \end{array} \right ),
 \eeq{sindwhw92936}
 which can be thought of as a section of the pullback of the tangent plus cotangent bundles (with    
   reversed
  parity), $X^*(\Pi (T \oplus T^*))$. The tensors in (\ref{definsdo339390}) may also be combined into a
   single object,
   \beq
    {\cal J} = \left ( \begin{array}{cc}
                               - J & P \\
                               L  & J^t 
                               \end{array} \right ) ,
   \eeq{singleajjawll}
 which is understood now  as ${\cal J} : T \oplus T^* \rightarrow T \oplus T^*$. 
  With this new notation we can rewrite (\ref{definsdo339390}) as follows
  \beq
   {\mathbf Q}_2 (\epsilon) = - \frac{1}{2} \int d\sigma d\theta\,\,\epsilon 
    \langle \Lambda, {\cal J} \Lambda  \rangle ,
  \eeq{dnelalw93889asl;}
   where $\langle\,\,,\,\,\rangle$ stands for the natural pairing defined in Appendix C.
   If such ${\mathbf Q}_2(\epsilon)$ obeying (\ref{djfkfllla;;;app292more}) exists then we will 
   say that there is an extended 
    supersymmetry on $\Pi T^*{\cal L}M$.  Indeed $\Pi T^* {\cal L}M$ admits an extended 
     supersymmetry if and only if $M$ is a generalized complex manifold with ${\cal J}$ being 
      a generalized complex structure
      \cite{Zabzine:2005qf} (for the basic definitions
      see Appendix C). The corresponding supersymmetry transformations are given by
\ber
\label{122333333333}&&\delta_2(\epsilon) \phi^\mu = \{ \phi^\mu, {\mathbf Q}_2 (\epsilon) \}=
i \epsilon D\phi^\nu J^\mu_{\,\,\nu} -  i \epsilon S_\nu P^{\mu\nu}, \\
\nonumber &&\delta_2(\epsilon) S_\mu =  \{ S_\mu, {\mathbf Q}_2 (\epsilon) \}=
i \epsilon D(S_\nu J^\nu_{\,\,\mu}) 
-\frac{i}{2} \epsilon S_\nu S_\rho P^{\nu\rho}_{\,\,\,\,,\mu}
 +i \epsilon D(D\phi^\nu L_{\mu\nu}) +\\
\label{122333333333A}&& \,\,\,\,\,\,\,\,\,\,\,\,\,\,\,\,\,\,\,\,\,\,\,\,\,\,\,\,\,\,\,\,\,\,\,\,\,\,+i \epsilon S_\nu D\phi^\rho J^\nu_{\,\,\rho,\mu} -
 \frac{i}{2} \epsilon D\phi^\nu D\phi^\rho L_{\nu\rho,\mu}.
\eer{bla920202002}
 If $\epsilon = \epsilon_0 + \theta \epsilon_1$ then the generator (\ref{dnelalw93889asl;}) 
 corresponds both to the supersymmetry generator (with  parameter $\epsilon_0$) 
  and to the generator of $U(1)$  R-symmetry (with parameter $\epsilon_1$).
The relation between an extended supersymmetry of the form  (\ref{122333333333}), (\ref{122333333333A}) and the interpretation of ${\cal J}$
as a generalized complex structure was first established in
\cite{Lindstrom:2004iw}.
  
  We may similarly consider the case with $H\neq 0$, where in the algebra (\ref{djfkfllla;;;app292more})
   the brackets $\{\,\,,\,\,\}$ are replaced by $\{\,\,,\,\,\}_H$ and ${\mathbf P}(a)$ is given by
    (\ref{definfoo38383830new}).  In this case an extended supersymmetry
     exists on $\Pi T^*{\cal L}M$ with $\{\,\,,\,\,\}_H$   if the manifold $M$ admits a twisted
      generalized complex structure (the Courant bracket is twisted by $H$,  see (\ref{deftwCour})).
       This result can be easily understood using the non-canonical transformation trick
        (\ref{perifofk83999}) and the property (\ref{noncanotr33782}) of the Courant bracket. Alternatively 
         we could just take the generator  ${\mathbf Q}_2(\epsilon)$ of the form (\ref{dnelalw93889asl;})
          and impose the algebra (\ref{djfkfllla;;;app292more}) with the Poisson bracket $\{\,\,,\,\,\}_H$ 
          and the correct ${\mathbf P}(a)$ (\ref{definfoo38383830new}).

\section{N=(2,2) supersymmetry in phase space}
\label{N2sigma}

In this Section we combine the Hamiltonians from Section \ref{N1sigma} with the extended supersymmetry
of  Section \ref{N2general}.
 In particular we are interested in the Hamiltonian picture of the $N=(2,2)$ sigma model 
  discussed in \cite{Gates:1984nk}. It means that in addition to the two supersymmetries 
  (\ref{transkd829292}) and (\ref{nonmanN1a})-(\ref{nonmanN1b})  we have to require the existence of two additional extended supersymmetries, ${\mathbf Q}_2$ and $\tilde{\mathbf Q}_2$. The different brackets
   between  the supersymmetry generators will give rise not only to the translation 
    generators in $\sigma$ but also in time, $t$. Thus the Hamiltonian (which is the generator
     of time translations) should appear in that context. Indeed we will reproduce 
      the full  $N=(2,2)$ algebra but in the Hamiltonian setup.
   
  As before we start from the case when $H=0$.  Using the notation of the previous Section we introduce two generators of the extended supersymmetry
\beq
  {\mathbf Q}_2 (\epsilon) = - \frac{1}{2} \int d\sigma d\theta\,\,\epsilon 
    \langle \Lambda, {\cal J}_1 \Lambda  \rangle,\,\,\,\,\,\,\,\,\,\,\,\,\,\,\,\,\,\,\,\,\,\,\,\,\,\,\,\,\,\,\,
      \tilde{\mathbf Q}_2 (\epsilon) = - \frac{1}{2} \int d\sigma d\theta\,\,\epsilon 
    \langle \Lambda, {\cal J}_2 \Lambda  \rangle . 
\eeq{twosneioehjal9}
 Since both generators satisfy the algebra (\ref{djfkfllla;;;app292more}) 
 ${\cal J}_1$ and ${\cal J}_2$ are generalized complex structures.  The bracket 
  between ${\mathbf Q}_2$ and $\tilde{\mathbf Q}_2$ should give rise to the generator of
   time translations, i.e. to the Hamiltonian (\ref{defhamilt})
   \beq
    \{ {\mathbf Q}_2(\epsilon), \tilde{\mathbf Q}_2(\tilde{\epsilon}) \} = {\cal H} (2i \epsilon \tilde{\epsilon}).
   \eeq{braytskwp2839}
The calculations of the bracket (\ref{braytskwp2839}) are tedious but straightforward.
  The final result can be summarized in the following relation between  the two generalized complex
   structures
\beq
 - {\cal J}_1 {\cal J}_2 = {\cal G} = \left (\begin{array}{cc}
                                                 0 & g^{-1}\\
                                                  g & 0 
                                                  \end{array} \right ),
\eeq{dhdkie930acnmdlas}
 where ${\cal G}$ is a generalized metric. From ${\cal G}^2 =1_{2d}$ it follows that 
  $[{\cal J}_1, {\cal J}_2]=0$. Thus the relation (\ref{braytskwp2839})
   forces $M$ to have a generalized K\"ahler structure. 

Alternatively, we could approach  the problem as follows: We introduce  a single extended
 supersymmetry generator ${\mathbf Q}_2$ and require that the Hamiltonian (\ref{defhamilt})
  is invariant under this supersymmetry, i.e. 
  \beq 
    \{ {\mathbf Q}_2(\epsilon), {\cal H}(a) \} =0.
  \eeq{indvahelle-2}
   The requirement (\ref{indvahelle-2}) leads to the algebraic condition
   \beq
    [{\cal J}_1, {\cal G}]=0~,
   \eeq{aldganksk389}
    where ${\cal G}$ is defined by the right hand side of (\ref{dhdkie930acnmdlas}), and to the differential conditions 
    \beq
     \nabla J_1=0,\,\,\,\,\,\,\,\,\,\,\,\,\,\,\,\,\nabla P_1=0,\,\,\,\,\,\,\,\,\,\,\,\,\,\,\,\,\nabla L_1=0,
    \eeq{dmdkllw9393888}
    where $\nabla$ is the Levi-Civita connection and $J_1, P_1, L_1$ are tensors defined through
     (\ref{singleajjawll}). Introducing ${\cal J}_2 = {\cal J}_1 {\cal G}$ it is straightforward to prove 
      that ${\cal J}_2$ is generalized complex structure and thus we have another extended supersymmetry $\tilde{\mathbf Q}_2$. Again we arrive at the conclusion that the invariance of 
       the Hamiltonian under the extended supersymmetry requires
         the generalized K\"ahler geometry. 
   Indeed using the Jacobi identity (\ref{Javoao20021})
   for $\{\,\,,\,\,\}$  (\ref{indvahelle-2}) follows from (\ref{braytskwp2839}). 
   
   The same line of reasoning can be applied to the case when $H\neq 0$.   The bracket 
    $\{\,\,,\,\,\}$ is then replaced by $\{\,\,,\,\,\}_H$ and the Hamiltonian ${\cal H}$ given by 
     (\ref{fullhamwithevetey}). The calculation of (\ref{braytskwp2839}) for $H\neq 0$ is the most demanding 
     calculation in this paper and it is gratifying that it yields the expected result; $M$ must carry a  twisted 
     generalized K\"ahler structure.

\section{Relation to the Gates-Hull-Ro\v{c}ek geometry}
\label{GHR}

 The geometry of the general $N=(2,2)$ sigma model is described in \cite{Gates:1984nk}.  
  The additional  supersymmetry transformations are given by
 \beq
  \delta (\epsilon^+, \epsilon^-) \Phi^\mu = i \epsilon^+ D_+ \Phi^\nu J^\mu_{+\nu} +  
 i  \epsilon^- D_- \Phi^\nu J^\mu_{-\nu},
 \eeq{transformGHR} 
  where $J_+$ and $J_-$ are the complex structures. Moreover the requirement that
  the transformations (\ref{transformGHR})  are symmetries of the action 
   (\ref{fullactionwthB}) leads to  the conditions
   \beq
    J^t_{\pm} g J_{\pm}=g,\,\,\,\,\,\,\,\,\,\,\,\,\,\,\,\,\,\,\,\,\,\,
    \nabla^{(\pm)} J_{\pm} =0,
   \eeq{conslal3900-3..}
    where $\nabla^{\pm}$ are the connections with torsion, $\Gamma^\pm = \Gamma \pm g^{-1}H$.
     This defines the bi-hermitian GHR geometry.  
     
     In the previous discussion we have provided a description of the same model  in 
      a Hamiltonian formulation. Thus the question arises: what is the relation between the Hamitonian     
      geometrical  data (two commuting (twisted) generalized complex structures) and the bi-hermitian 
       GHR geometry?  Surprisingly the answer to this question is very easy to obtain. 
        Namely we have to take the transformation (\ref{transformGHR}) and reduce it to the Hamiltonian 
         superfields $\phi$ and $S$, (\ref{newsu3902}) and then to compare the result of the reduction 
          to the transformations (\ref{122333333333}) and (\ref{122333333333A}).  From this comparison 
          we find the relation between ${\cal J}_{1}, {\cal J}_{2}$
           and $(J_\pm, g)$. 
       
   We first illustrate this by writing the transformation of $\phi$ that follows from
  (\ref{newsu3902}) 
$$  \delta \phi^\mu =\frac{1}{2} i \epsilon_1 D_1 \phi^\nu (J^\mu_{+\nu} + J^\mu_{-\nu}) +
  \frac{1}{2}i \epsilon_2 D_1 \phi^\nu (J^\mu_{+\nu} - J^\mu_{-\nu}) -$$
  \beq
  \frac{1}{2} i\epsilon_1 S_\nu ((\omega_+^{-1})^{\mu\nu} - (\omega_-^{-1})^{\mu\nu})
   - \frac{1}{2}i\epsilon_2 S_\nu ((\omega_+^{-1})^{\mu\nu} + (\omega_-^{-1})^{\mu\nu}) ,
 \eeq{newphishjat}
 where   $\epsilon_1\equiv \frac{1}{\sqrt{2}}(\epsilon^+ + i \epsilon^-)$ and 
 $\epsilon_2\equiv  \frac{1}{\sqrt{2}}(\epsilon^+ - i \epsilon^-)$.  The expression (\ref{newphishjat})
  should be compared to (\ref{122333333333}), and then do the same with the transformations
   for $S$.  
   Since the expression for this transformation is quite lengthy while the calculations 
    are straightforward\footnote{In the case with $S$ there will appear a term with $\d_0 \phi$
     which we can replace 
    by the corresponding Hamiltonian equations of motion. 
     This is not surprising since the transformations (\ref{transformGHR}) close 
     on-shell while the transformations (\ref{122333333333})-(\ref{122333333333A}) 
     close off-shell.}, we just give the final result of the identifications
\beq
 {\cal J}_{1,2} = - \frac{1}{2} 
 \left ( \begin{array}{ll}
       J_+ \pm J_- & -(\omega_+^{-1} \mp \omega_-^{-1}) \\
     \omega_+ \mp \omega_- & - (J_+^t \pm J^t_-)
\end{array} \right )
\eeq{degindoftwocvom} 
 This is exactly the same as in Gualtieri's thesis \cite{gualtieri}, modulo an insignificant overall sign.
 
 Thus this Section concludes a physical derivation of (twisted) generalized K\"ahler geometry
  and its identification with the bi-hermitian GHR geometry. As matter of fact  the (twisted) generalized 
   K\"ahler geometry is the Hamiltonian reformulation of bi-hermitian GHR geometry. 
   
   Alternatively we may state this as follows: Just like in the Lagrangian formulation the $N=(1,1)$ sigma 
   model has a second supersymmetry iff the target space has GHR geometry, in the Hamiltonian formulation
   the $N=(1,1)$ sigma model has a second supersymmetry iff the target space has (twisted) GK geometry.
   
   We stress that although it was previously known that GHRG can be mapped to GKG, GHRG is derived from a 
   sigma model with only on-shell closure of the algebra (in general) and target tangent space fields only.
   Here we have discussed a model with off-shell closure and both target tangent space and cotangent space fields.
   This has allowed us to derive GKG, as well as the mapping to GHRG, directly from the sigma model.
  
\section{Topological twist in phase space}
\label{twist}

 In this Section we use our picture of $N=(2,2)$ sigma model to discuss
  topological twist and the corresponding topological field theories (TFT). Without much 
   effort we will identify the twisted $N=(2,2)$ sigma model with the TFTs previously discussed
    in \cite{Alekseev:2004np} and \cite{Bonelli:2005ti}. 

As pointed out in \cite{Zabzine:2005qf} the extended supersymmetry transformations
   (\ref{122333333333}),  (\ref{122333333333A})  can be twisted and there exists associated BRST transformation. The supersymmetry generators (\ref{manidkalalal122}) 
    and (\ref{definsdo339390}) can be converted to odd generators
   by formally setting the odd parameter $\epsilon$ to one. However these generators will still satisfy 
    the same algebra. Thus we can define an odd generator
    \beq
    {\mathbf q} = {\mathbf Q}_1(1) + i {\mathbf Q}_2 (1)= - \int d\sigma d\theta\,\, (S_\mu Q \phi^\mu +
     i D\phi^\nu S_\mu J^\nu_{\,\,\,\mu}  + \frac{i}{2} D\phi^\mu D\phi^\nu L_{\mu\nu} + \frac{i}{2}
      S_\mu S_\nu P^{\mu\nu} ),
    \eeq{definsoaow73930}
 which is nilpotent due to the algebra of supersymmetry generators ${\mathbf Q}_1$ and 
  ${\mathbf Q}_2$. The generator ${\mathbf q}$ generates the following transformations
\ber
\label{blalalls---}&&{\mathbf s} \phi^\mu = \{ {\mathbf q}, \phi^\mu \} =  
 -iQ\phi^\mu - J^\mu_{\,\,\,\nu} D\phi^\nu
 + P^{\mu\nu} S_\nu, \\
\nonumber &&{\mathbf s} S_\mu = \{ {\mathbf q}, S_\mu \} = - i Q S_\mu - D(S_\nu J^\nu_{\,\,\,\mu})
 + \frac{1}{2} S_\nu S_\rho P^{\nu\rho}_{\,\,\,\,\,\,\mu} - D(D\phi^\nu L_{\mu\nu}) -\\
\label{blalsjswodmxmx} &&\,\,\,\,\,\,\,\,\,\,\,\,\,\,\,\,\,\,\,\,\,\,\,\,\,\,\,\,\,\,\,- S_\nu  D\phi^\rho
  J^\nu_{\,\,\,\rho,\mu} + \frac{1}{2} D\phi^\nu D\phi^\rho L_{\nu\rho,\mu},
\eer{blavla}
 such that ${\mathbf s}^2=0$. Therefore
as an alternative  we can relate the generalized complex structure on $M$ with an odd differential 
 ${\mathbf s}$ on $C^\infty (\Pi T^* {\cal L}M)$.  In the same way for $H\neq 0$ we can introduce ${\mathbf s}$ on $C^\infty (\Pi T^* {\cal L}M)$ with $\{\,\,,\,\,\}_H$ instead of $\{\,\,,\,\,\}$.
   The generator (\ref{definsoaow73930})  is reminiscent of the solution of the master equation
    proposed in \cite{Zucchini:2004ta}. However there are
   conceptual differences in the setup and in the definitions
     of the basic operations (e.g., $D$).
  
  Next we consider our $N=(2,2)$ sigma model in our Hamiltonian setup as described in previous 
   sections.  We consider the case with $H\neq 0$.
   Since we have two complex structures ${\cal J}_1$ and $ {\cal J}_2$ there are two 
    BRST transformations we can introduce, ${\mathbf s}_1$ and ${\mathbf s}_2$.  For the sake of 
     concreteness we choose the generalized complex structure ${\cal J}_1$ and the corresponding 
      BRST transformation ${\mathbf s}_1$.  Then due to the relation (\ref{braytskwp2839}) we can 
      write  the action  (\ref{hamsle03440})  as follows
  \beq    
  S = \int d^2\sigma\,d\theta\,\,\left (i (S_\mu - B_{\mu\nu} D\phi^\nu) \d_0 \phi^\mu  +\frac{i}{2}
   {\mathbf s}_1 
   \tilde{\mathbf Q}_2(1) \right ).
 \eeq{Plocalizacaskk1}
  Thus the Hamiltonian ${\cal H}$ (\ref{fullhamwithevetey}) is BRST exact. Due to the standard 
   argument (e.g., see \cite{Witten:1991zz}) 
    the theory will be localized on the fixed points of action of BRST 
    transformations. If we expand BRST transformations (\ref{blalalls---}), (\ref{blalsjswodmxmx}) in components and look
     at purely bosonic fixed points (i.e., $\lambda=0$ and $\rho =0$), we arrive
 at the following description of them    
 \beq
  \frac{1}{2} (1_{2d} + i {\cal J}_1 ) \left ( \begin{array}{c}
                                           \d X\\
                                            p 
                                            \end{array} \right ) = 0 ,
 \eeq{duaksllld002-;}
  which is equivalent to say that there is a set of first class constraints 
  \beq
   v^\mu p_\mu + \xi_\mu  \d X^\mu =0,
  \eeq{constarsuiw73492}
   where $(v+ \xi)$ is the section of the subbundle  $L$ associated with the decomposition 
    $T \oplus T^* = L \oplus \bar{L}$ given by a (twisted) generalized complex structure ${\cal J}_1$. 
     The theory which is defined by the constraint (\ref{constarsuiw73492}) was originally discussed
      in \cite{Alekseev:2004np}.  These theories were further discussed in \cite{Bonelli:2005ti} and 
      the standard A- and B-models fit this description.
      Therefore we  conclude that  the action   (\ref{Plocalizacaskk1}) is the gauged 
       fixed action for the theory defined by (\ref{constarsuiw73492}).
          The (twisted) generalized complex structure 
        ${\cal J}_1$ defines the TFT while another structure ${\cal J}_2= {\cal J}_1 {\cal G}$ is 
         used for the gauge fixing.  Alternatively we may say that we use the generalized metric 
          ${\cal G}$ for the gauge fixing.  The first term in the action (\ref{Plocalizacaskk1}) 
          \beq
           \int \Theta = i \int dt d\sigma d\theta \,\, (S_\mu - B_{\mu\nu} D\phi^\nu) \d_0 \phi^\mu
          \eeq{lidosuakkk}
 with $\Theta$ as given in (\ref{indelivsis999more})  can be interpreted as a topological term
 and may be analyzed by the same methods as in
  \cite{Bonechi:2005mw}.  Indeed our considerations are in complete agreement with the previous 
   discussions   \cite{Kapusti2003sg},  \cite{Kapustin:2004gv}  of topological 
    twist of general $N=(2,2)$ sigma model. The condition (\ref{duaksllld002-;})  can be interpreted
     as the phase space description of a generalized instanton \cite{Kapusti2003sg}.

  Alternatively we can exchange the role of ${\cal J}_1$ and ${\cal J}_2$, i.e. the BRST
    transformation ${\mathbf s}_2$ is defined by ${\cal J}_2$ and the gauge fixing is given 
     by ${\cal J}_1$ (or ${\cal G}$). The action is given now
     \beq    
  S = \int d^2\sigma\,d\theta\,\,\left (i (S_\mu - B_{\mu\nu} D\phi^\nu) \d_0 \phi^\mu  +\frac{i}{2}
   {\mathbf s}_2 
   {\mathbf Q}_2(1) \right ).
 \eeq{Plocalizacaskk1new}
 Obviously we can repeat the previous  analysis and this model presents the alternative 
  topological twist of sigma model.  As expected the topological twist of $N=(2,2)$ sigma model 
   gives rise to two non equivalent TFTs associated with two different (twisted)
    generalized complex structures.

\section{Summary}
\label{end}

We have presented a physical derivation of Gualteri's result \cite{gualtieri}
 on the equivalence between the (twisted) generalized K\"ahler geometry and 
  Gates-Hull-Ro\v{c}ek geometry. We explained this equivalence in terms of $N=(2,2)$
   supersymmetric sigma models and in this language these two geometrical descriptions
    correspond to the Hamiltonian and the Lagrangian formalisms, respectively.    
  As an application of our results we have briefly discussed  the topological twist 
   of $N=(2,2)$ sigma models in the Hamiltonian setup. The formalism allows us to identify 
    the corresponding TFTs.
We believe that the presented
     Hamiltonian formalism has potential for further applications.
     Hopefully, it might shed  light on some of the issues of mirror symmetry, in particular
      in the context of generalized geometry. 

\noindent{\bf Acknowledgement}: 
The research of U.L. is supported by VR grant 650-1998368.  
The research of M.Z. is supported by  by VR-grant 621-2004-3177.
Partial support for this research is further provided  by EU grant (Superstring theory)
MRTN-2004-512194.

\appendix
\Section{Appendix: $N=(1,1)$ supersymmetry}
\label{a:11susy}

In this appendix we collect our notation for $N=(1,1)$  
superspace. 

We use real (Majorana) two-component spinors $\psi^\alpha=
(\psi^+, \psi^-)$. Spinor indices are raised and lowered with the  
second-rank antisymmetric symbol $C_{\alpha\beta}$, 
which defines the spinor inner product:
\beq
C_{\alpha\beta}=-C_{\beta\alpha}=-C^{\alpha\beta}~,\qquad C_{+-} 
=i~,\qquad
\psi_\alpha =\psi^\beta C_{\beta\alpha}~,\qquad \psi^\alpha= C^ 
{\alpha\beta} \psi_\beta~.
\eeq{Cdef}
Throughout the paper we use  $(\+,=)$ as worldsheet indices, and $ 
(+,-)$ as two-dimensional spinor
indices.  We also use superspace conventions where the pair of spinor
coordinates of the two-dimensional superspace are labelled $\theta^{\pm}$,
and the spinor derivatives $D_\pm$ and supersymmetry generators
$Q_\pm$ satisfy
\ber
D^2_+ &=&i\d_\+~, \qquad
D^2_- =i\d_=~, \qquad \{D_+,D_-\}=0~,\cr
Q_\pm &=& iD_\pm+ 2\theta^{\pm}\d_{\pp}~,
\eer{alg}
where $\d_{\pp}=\partial_0\pm\partial_1$. 
The supersymmetry transformation of a superfield $\P$ is given by
\ber
\delta \P &\equiv &-i(\e^+Q_++\e^-Q_-)\P \cr
&=& (\e^+D_++\e^-D_-)\P
-2i(\e^+\theta^+\d_\++\e^-\theta^-\d_=)\P ~.
\eer{tfs}
The components of a scalar superfield $\P$ are defined by  
projection as follows:
\ber
\P|\equiv X~, \qquad D_\pm\P| \equiv \p_\pm~, \qquad D_+D_-\P|\equiv F~,
\eer{comp}
where the vertical bar $|$ denotes ``the $\theta =0$ part''.
The $N=(1,1)$  spinorial measure is conveniently written in terms of spinor derivatives:
\beq
\left.\int d^2\theta \,\,{\cal L} =   (D_+ D_- {\cal L})\right| .
\eeq{sssssssssspp}

\Section{Appendix: superPoisson algebra}
\label{a:superpoisson}

In this Appendix we collect the conventions on the supersymmetric version of Poisson 
 brackets. 

As phase space we consider the cotangent bundle $\Pi T^*{\cal L}M$ of the superloop 
 space ${\cal L}M= \{\phi : S^{1,1} \rightarrow M\}$.
 The canonical symplectic structure is given by
\beq
 \omega = i \int d\sigma d\theta\,\, \delta S_\mu \wedge \delta \phi^\mu
\eeq{definalsdp}
 and thus the cotangent bundle $\Pi T^*{\cal L}M$ has a reversed parity on the fibers 
  ($\Pi$ is to remind us of the reversed parity).
  The  symplectic structure (\ref{definalsdp}) makes $C^{\infty}(\Pi T^*{\cal L}M)$ (the space of smooth 
   functionals on $\Pi T^*{\cal L}M$) into a superPoisson algebra.  The space $C^{\infty}(\Pi T^*{\cal L}M)$
    has a natural $\mathbb{Z}_2$ grading $|\cdot |$, with $|F|=0$ for even and $|F|=1$ for odd functionals. 
 
 For a functional $F(S, \phi)$ we define the left and right functional derivatives 
  as follows
 \beq
 \delta F = \int d\sigma d\theta \left ( \frac{ F \overleftarrow{\delta}}{\delta S_\mu} \delta S^\mu +
  \frac{F \overleftarrow{\delta}}{\delta \phi^\mu} \delta \phi^\mu \right ) =
  \int d\sigma d\theta \left ( \delta S_\mu \frac{ \overrightarrow{\delta} F}{\delta S_\mu}  +
  \delta \phi^\mu \frac{\overrightarrow{\delta} F}{\delta \phi^\mu}  \right ) .
 \eeq{Pdefinals;a[[[}
  Using this definition, the Poisson bracket corresponding to (\ref{definalsdp}) is given by
  \beq
  \{ F, G\} = i \int d\sigma d\theta \left ( \frac{ F \overleftarrow{\delta}}{\delta S_\mu} \frac{\overrightarrow{\delta} G}{\delta \phi^\mu} - \frac{F \overleftarrow{\delta}}{\delta \phi^\mu} \frac{\overrightarrow{\delta} G}
  {\delta S_\mu} \right ). 
  \eeq{definape8349}
   This bracket satisfies the appropriate graded versions of antisymmetry, of the Leibnitz rule and 
    of the Jacobi identity
   \beq
    \{ F, G\} = - (-1)^{|F| |G|} \{G, F\},
  \eeq{ansiruwiqw99}
  \beq
   \{ F, GH \} = \{ F, G\} H + (-1)^{|F| |G|} G \{ F, H\},
  \eeq{Leibniziao2290}
  \beq
  (-1)^{|H||F|} \{ F, \{G, H\}\} + (-1)^{|F| |G|} \{ G, \{ H, F\}\} + (-1)^{|G| |H|} \{ H, \{ F, G \} \} =0.
  \eeq{Javoao20021}
  
  Indeed there is whole family of symplectic structures on $\Pi T^*{\cal L}M$ parametrized by 
   a closed three form $H$
     \beq
  \omega = i \int d\sigma d\theta\,\, \left ( \delta S_\mu \wedge \delta \phi^\mu - H_{\mu\nu\rho}
   D\phi^\mu\, \delta \phi^\nu \wedge \delta \phi^\rho \right ),
 \eeq{definsymlpedjsdomore}
 with the Poisson bracket defined by
  \beq
  \{ F, G\}_H = i \int d\sigma d\theta \left ( \frac{ F \overleftarrow{\delta}}{\delta S_\mu} \frac{\overrightarrow{\delta} G}{\delta \phi^\mu} - \frac{F \overleftarrow{\delta}}{\delta \phi^\mu} \frac{\overrightarrow{\delta} G}
  {\delta S_\mu}  + 2  \frac{ F \overleftarrow{\delta}}{\delta  S_\nu} H_{\mu\nu\rho} D\phi^\mu 
   \frac{\overrightarrow{\delta} G}{\delta S_\rho} \right  ). 
  \eeq{definape8349moreysj}
  It is straightforward to show that $C^{\infty} (\Pi T^* {\cal L}M)$ with $\{\,\,,\,\,\}_H$ is a superPoisson algebra, 
   i.e. that it satisfies  (\ref{ansiruwiqw99})-(\ref{Javoao20021}).

\Section{Appendix: generalized complex geometry}
\label{TplusTstar}

Consider the vector bundle $T \oplus T^*$ which is the sum of the tangent and cotangent 
 bundles of an $d$-dimensional manifold $M$. $T\oplus T^*$ has a natural pairing
\beq
  \langle v + \xi, w + \eta \rangle \equiv \frac{1}{2}(i_w \xi + i_v \eta) \equiv\frac{1}{2}
 \left (\begin{array}{l}
 v\\
 \xi \end{array} \right )^t {\cal I}
\left (\begin{array}{l}
 w\\
 \eta \end{array}\right ) ,
\eeq{definnaturka}
 where $(v + \xi), (w + \eta) \in T\oplus T^*$.
 The smooth sections of $T\oplus T^*$ have a natural bracket operation called
  the Courant bracket and defined as follows
\beq
 [v+ \xi, w + \eta]_c = [v, w] + {\cal L}_v \eta - {\cal L}_w \xi  -\frac{1}{2}d (i_v\eta - i_w \xi) ,
\eeq{defCourant}
 where $[\,\,,\,\,]$ is a Lie bracket on $T$. Given a closed three form $H$ we can define
  a twisted Courant bracket
\beq
 [ v+ \xi, w + \eta]_H = [v+\xi, w+\eta ]_c + i_v i_w H .
\eeq{deftwCour}
 The orthogonal automorphism (i.e., such which preserves $\langle\,\,,\,\,\rangle$)
   $F:T\oplus T^* \rightarrow T\oplus T^*$ of the (twisted)
 Courant bracket 
\beq
 F([v+\xi, w + \eta ]_H) = [F(v+\xi), F(w+\eta)]_H
\eeq{automspaorp}
 is a semidirect product of $Diff(M)$ and $\Omega^2_{closed}(M)$,
 where the action of the closed two form is given by
\beq
 e^b(v+\xi) \equiv v + \xi + i_v b
\eeq{actionwtalk}
 for $b\in \Omega^2_{closed}(M)$. The transformation (\ref{actionwtalk}) is called a $b$-transform.
  If $b$ is a non-closed two form then the following holds
  \beq
   [ e^b(v+ \xi), e^b(w + \eta)]_c = e^b[v+\xi, w+\eta ]_c + i_v i_w db . 
  \eeq{noncanotr33782}
 A maximally isotropic subbundle $L$ of    $T\oplus T^*$, which is involutive with respect 
 to the (twisted) Courant bracket is called a (twisted) Dirac structure. We can consider two complementary 
 (twisted) Dirac structures $L_+$ and $L_-$ such that $T\oplus T^* = L_+ \oplus L_-$. Alternatively 
 we can define $L_\pm$ by providing a map ${\cal J}:T\oplus T^* \rightarrow T\oplus T^*$ with 
 the following properties
\beq
 {\cal J}^t {\cal I} = - {\cal I} {\cal J},\,\,\,\,\,\,\,\,
 {\cal J}^2 = 1_{2d},\,\,\,\,\,\,\,\,
 \Pi_{\mp} [\Pi_{\pm} (v+\xi), \Pi_{\pm}(w+\eta)]_H =0
\eeq{transverds}
 where $\Pi_\pm=\frac{1}{2} (1_{2d} \pm {\cal J})$ are projectors on $L_\pm$.

 A (twisted) generalized complex structure is the complex version of two complementary (twisted) 
 Dirac subbundles
 such that  $(T\oplus T^*)\otimes {\mathbb C} = L \oplus \bar{L}$. We can define the generalized 
 complex structure as a map ${\cal J}: (T\oplus T^*)\otimes {\mathbb C} 
  \rightarrow (T\oplus T^*)\otimes {\mathbb C}$ with 
 the following properties
\beq
 {\cal J}^t {\cal I} = - {\cal I} {\cal J},\,\,\,\,\,\,\,\,
 {\cal J}^2 = - 1_{2d},\,\,\,\,\,\,\,\,
 \Pi_{\mp} [\Pi_{\pm} (v+\xi), \Pi_{\pm}(w+\eta)]_H =0,
\eeq{transverds}
 where $\Pi_\pm=\frac{1}{2} (1_{2d} \pm i{\cal J})$ are the projectors on $L$ and $\bar{L}$ correspondingly. 
 
  We can define a generalized metric on $T\oplus T^*$ as  a subbundle $C_+$ of $T\oplus T^*$
   of rank $d$ such that the induced metric (from the natural pairing on $T\oplus T^*$) on $C_+$
    is positive definite. 
      Alternatively we may define the generalized metric as a map ${\cal G} : T\oplus T^* \rightarrow
    T\oplus T^*$ such that ${\cal G}^2=1_{2d}$ and ${\cal G}^t {\cal I} = {\cal I} {\cal G}$. Thus 
     $\frac{1}{2}(1_{2d}+ G)$ is a projector to $C_+$. 
 
 A (twisted) generalized K\"ahler structure is a pair $({\cal J}_1, {\cal J}_2)$ of commuting (twisted)     
 generalized  complex structures such that ${\cal G} = -{\cal J}_1 {\cal J}_2$ is a positive definite
  generalized metric on $T\oplus T^*$. 

For further details the reader may consult  Gualtieri's thesis \cite{gualtieri}.

\end{document}